\documentstyle[12pt,epsfig]{article}

\begin{document}
\begin{flushright}
 Preprint IHEP-96-82\\
 hep-ex/0110011

\end{flushright}
\vskip 0.5cm
\vskip 1in
\centerline{ \bf\Large {Production Asymmetry Measurement 
of High $x_{T}$ Hadrons}}

\centerline{ \bf\Large {in $p{\uparrow}p$-Collisions at 40 GeV} }
\vskip 0.01in
\begin{center}
V.V.Abramov$^{1}$, A.S.Dyshkant$^{1}$, 
V.N.Evdokimov$^{1}$, P.I.Goncharov$^{1}$,
A.M.Gorin$^{1}$, A.N.Gurzhiev$^{1}$,  
Yu.P.Korneev$^{1}$, A.V.Kostritskii$^{1}$, 
A.N.Krinitsyn$^{1}$, V.I.Kryshkin$^{1}$,  
Yu.M.Mel'nik$^{1}$, V.M.Podstavkov$^{1}$, 
 N.I.Sin'kin$^{2}$, S.I.Tereshenko$^{1}$, 
 L.K.Turchanovich$^{1}$, A.E.Yakutin$^{1}$, 
 A.A.Zaichenko$^{1}$, V.N.Zapol'sky$^{1}$
\end{center}
\begin{center}
$^{1}$Institute for High Energy Physics, Protvino,  Russia %
\end{center} 
\begin{center}
$^{2}$Institute of Theoretical and Experimental Physics, Moscow, Russia %
\end{center}
\begin{abstract}
Single-spin
asymmetries for hadrons have been measured in collisions of
transversely-polarized
40 GeV/c proton beam  with an unpolarized liquid hydrogen target. The
asymmetries were measured for $\pi^{\pm}$, $K^{\pm}$, protons and 
antiprotons, produced in the central region ($0.02 \le x_{F} \le 0.10$ and
$0.7 \le p_{T} \le 3.4$ GeV/c). Asymmetries for $\pi^{\pm}$,
$K^{\pm}$ and  \=p show within measurement errors the linear dependence on
$x_{T}$ and change a sign near 0.37. For protons negative asymmetry, 
independent of $x_{T}$ has been found. The results are compared with those 
of other experiments and SU(6) model predictions.
\end{abstract}
\vskip 1.0cm

\centerline{Submitted to {\em Nucl. Phys. B\/} }
\newpage

\section{Introduction}

  It is widely believed that the role of spin is decreasing with
the rise of beam energy and transverse momentum ($p_{T}$) of secondary
particles. In particular, the perturbative QCD predicts
the vanishing of spin effects
with the rise of $p_{T}$~\cite{SP2}. 
Experiments performed during last  years
have shown much more complicated picture. It was found that polarization
of \mbox{$\Lambda$-hyperons} produced in hadron-hadron 
collisions is almost energy
independent and  is rising with $p_{T}$ and $x_{F}=2p^{*}_{Z}/\sqrt{s}$,
where $p^{*}_{Z}$ and $\sqrt{s}$ are, respectively, longitudinal momentum
of secondary particle and total reaction energy in C.M. \cite{SP3}.
Single-spin asymmetry ($A_{N}$) is rising with $p_{T}$ 
and $x_{F}$  at least for $\pi$- mesons and it  differs from zero
up to the highest available 
beam energies (200 GeV) ~\cite{SAR},~\cite{E704XF}.
 
  It seems, that the most promising among theoretical approaches 
to the solution  of large single-spin asymmetry problem, 
observed in experiments, 
is to take into account higher twists contributions (see, for
example~\cite{EFR95}).

  The existing data on single-spin asymmetries have in most cases, limited
statistical accuracy, $x_{T}=2p_{T}/\sqrt{s}$
and $x_{F}$ ranges  as well as incomplete particle
identification. That does not allow one to make definite conclusions about
asymmetry dependence on kinematical variables and quantum numbers of
hadrons. 

The goal of the 
experiment is to study single-spin asymmetry ($A_{N}$) in reactions:
\begin{equation}
 p\uparrow + p(A) \rightarrow h + X, \quad
\label{REAC}
\end{equation}
where
$h$ denotes $\pi^{\pm}$, $K^{\pm}$, p, \=p or hadron pair~\cite{BEAM}.

It includes, in particular, the following points:
\begin{itemize} 
\item {$A_{N}$ vs $x_{T}$ and $x_{F}$ ($x_{T}$ and $x_{F}$ scaling check);}
\item {$A_{N}$ vs quark flavor (u,d,s);}
\item {$A_{N}$ vs A - atomic number (nature of A-dependence);}
\item {$A_{N}$ vs $x_{T}$ for hadron pairs.}
\end{itemize} 

  The experiment has been carried out with
FODS-2 double arm spectrometer designed to
study high $p_{T}$ single and pair hadron production in pp-, $\pi$p-, pA- and
$\pi$A- collisions~\cite{FODS}.  
A new IHEP polarized 40 GeV/c proton 
beam gives an additional tool to study high $p_{T}$ processes in
reactions (\ref{REAC}).
  The beam energy and available intensity of polarized beam allow one to
measure $A_{N}$ in a wide $x_{F}$ and $x_{T}$ range ($x_{T}<0.9$)
for not so high $p_{T}$ (up to 3.4 GeV/c) of charged hadrons.

  Preliminary results of this experiment, based on half the statistics,
were reported at SPIN-95 workshop in Protvino~\cite{SPIN95}.
\section {Beam}
Polarized protons~\cite{BEAM} are produced
via decays of  $\Lambda$-hyperons. A slow extracted 70~GeV/c proton beam
with intensity up to $10^{13}$ ppp strikes  the primary target.
Charged particles, produced in the primary target,
are deflected and absorbed in a beam dump. Protons from $\Lambda$-decays
 are collimated in vertical direction to select polarization sign
and beam intensity.
Average polarization of 40 GeV/c beam
(for $\Delta p/p=\pm4.5\%$) is $39_{-3}^{+1}\%$ and it was estimated using 
Monte Carlo simulation of the polarized beam production.
The vertical beam position on the secondary target is
corrected by two magnets. The beam
polarization sign  changes automatically  each 15 minutes
during 3-4 spills. Beam monitoring is performed
by ionization chambers and scintillator monitors (arm triggers) with relative 
precision better than 1\%. Calculated and measured
beam parameters are in a good agreement, which gives us confidence in the
knowledge of its polarization calculation.
\section{Spectrometer}
The experimental setup \mbox{FODS-2} is a rotating double arm 
spectrometer ~\cite{FODS} shown schematically in Fig.1.
It includes
a two gap magnet, drift chambers to measure momentum and
angles of particles production,
two Cerenkov ring imaging detectors (SCOCH) to identify
particles, scintillation counters, hodoscopes and hadronic calorimeters to 
make trigger decision. The SCOCH detector is able to identify hadrons in the
following momentum ($p$) ranges:
$\pi^{\pm}$ 
($2 < p < 24$ GeV/c), 
$K^{\pm}$ ($4 < p < 24$ GeV/c), p and \=p from 8
up to 24 GeV/c. Protons and antiprotons were also identified in
the momentum range from 5 to 8 GeV/c using the absence of a signal in the
SCOCH detector.
\noindent
The transverse position ($X,Y$) of an interaction point in the hydrogen
target ($D$=7~cm, $L$=40~cm, $L/\lambda_{abs}$=5\%)
is measured with
beam hodoscopes.
Longitudinal coordinate ($Z$) is determined as a result of matching 
$X,Y$-coordinates, measured with beam hodoscopes, with a particle trajectory,
reconstructed using drift chambers downstream of the magnet.
 Hodoscopes resolution is
$\pm$1 mm, both for $X$ and $Y$ coordinates. 
The total number of channels in each
hodoscope plane is 32, and average multiplicity was around 2~per plane. When
the multiplicity is not equal to unity a hodoscope 
channel closest to the center of
gravity for the beam during the spill is used. Ionization chambers were
used to determine the center of gravity for the beam.
\section {Measurements}
   The measurements were carried out during 6 days
of the November Run in 1994. The mean intensity of the polarized
beam was $9 \cdot 10^{6}$~ppp,
limited by radiation environment. Now
the shielding has been improved and nominal intensity
up to $2.6 \cdot 10^7$ can be achieved.
 The trigger for each arm required
signals from the scintillation counters and energy deposition above threshold
in the hadronic calorimeter. The total number of 
polarized protons that striked the target was about $1.8\cdot 10^{11}$
for each polarization sign. The corresponding number of recorded events
is $6.3 \cdot 10^{5}$/polarization/arm. 
24\% of them have been reconstructed and 
identified for the final analysis.
\section {Data reduction}

   The reconstructed trajectory of a particle downstream  the spectrometer 
magnet is used to determine its momentum and production angles.
Since the magnet deflects charged particles in vertical direction it does
not change essentially polar angle ($\theta$), but changes azimuthal
angle ($\phi$) depending on momentum and sign of the particle. The 
mean value of $cos{\phi}$ varies from 0.80 to 0.89.

  Particle momentum was determined by matching incident particle transverse
coordinates in the target ($X,Y$), measured with the beam hodoscopes
having the trajectory downstream the magnet. The  momentum ($p$) of hadron 
is used to determine its mass ($m$):
\begin{equation}
m^2 = p^{2}(1-\beta^{2})/\beta^{2}, \quad 
\end{equation}
where $\beta$ is hadron velocity, 
measured with the SCOCH detector~\cite{SCOCH}.
Mass spectrum ($m^2$) of hadrons measured with the SCOCH is shown in
Fig.2. For heavy hadrons (protons) the main error in  $m^2$ is due to
the uncertainty in the momentum. For light particles ($\pi$-meson) the
error in $m^2$ is mainly due to the uncertainty~in~$\beta$.
\section {Results}
 Single-spin asymmetry has been calculated for 
two arms ($L$-left, $R$-right):
\begin{equation}
 A^{L}_{N} = { 1\over{P_{B}\cdot cos{\phi}} } \cdot 
{ {N^{u}_{L}-N^{d}_{L}} \over {N^{u}_{L}+N^{d}_{L}} },   \quad 
\end{equation}
\begin{equation}
 A^{R}_{N} = {-1\over{P_{B}\cdot cos{\phi}}} \cdot 
{ {N^{u}_{R}-N^{d}_{R}} \over {N^{u}_{R}+N^{d}_{R}} },  \quad 
\end{equation}
where $P_{B}$ is the average beam polarization (39\%), 
and $N^{u(d)}_{L(R)}$ is
the  number of normalized events, for protons,
 polarized up ($u$) or down ($d$) for the 
left ($L$) or right ($R$) arm. The so called canonical 
helicity frame is used here
for asymmetry definition, as it is in ~\cite{SAR}.
Asymmetries for two arms have been averaged, which decreased
some systematic errors and doubled the statistics. Only statistical errors are
taken into account in the  results. The relative systematic error
in $A_{N}$ due to possible errors in beam polarization, beam monitoring
and $cos{\phi}$ measurement is estimated to be less than 20\%.

Along with the asymmetries for
single particles the asymmetries for particle ratios have been calculated
as well since they are free from beam monitoring problems.

  Asymmetries for $\pi^{\pm}$, $K^{\pm}$, protons and antiprotons are
presented as functions of $p_T$ in Table~\ref{Tab1},~\ref{Tab2} and
\ref{Tab3}, respectively.
\begin{table}[htb]
\small
\caption{$A_N$ vs $p_T$ for $\pi^+$ and $\pi^-$ -mesons.}
\begin{center}
\begin{tabular}{|c|c|c|c|c|}    \hline
             &                   &                    \\[0.01cm] 
$p_T$ (GeV/c)& $A_{N}^{\pi^{+}}$ & $A_{N}^{\pi^{-}}$  \\[0.3cm] \hline
 0.661 & -0.12 $\pm$ 0.10  & -0.017$\pm$ 0.080 \\[0.3cm] \hline
 0.912 & -0.085$\pm$ 0.046 & -0.040$\pm$ 0.033 \\[0.3cm] \hline
 1.131 & -0.018$\pm$ 0.019 & -0.005$\pm$ 0.012 \\[0.3cm] \hline
 1.368 & -0.027$\pm$ 0.013 & -0.020$\pm$ 0.011 \\[0.3cm] \hline
 1.613 & +0.009$\pm$ 0.014 & -0.037$\pm$ 0.014 \\[0.3cm] \hline
 1.861 & +0.002$\pm$ 0.019 & -0.018$\pm$ 0.022 \\[0.3cm] \hline
 2.110 & +0.035$\pm$ 0.027 & -0.027$\pm$ 0.035 \\[0.3cm] \hline
 2.360 & +0.091$\pm$ 0.039 & +0.068$\pm$ 0.056 \\[0.3cm] \hline
 2.613 & +0.065$\pm$ 0.058 & -0.062$\pm$ 0.081 \\[0.3cm] \hline
 2.864 & -0.012$\pm$ 0.083 & -0.05 $\pm$ 0.11  \\[0.3cm] \hline
 3.118 & +0.20 $\pm$ 0.11  & -0.24 $\pm$ 0.16  \\[0.3cm] \hline
 3.368 & +0.21 $\pm$ 0.16  & -0.21 $\pm$ 0.20  \\[0.3cm] \hline
\end{tabular}
\end{center}
\label{Tab1}
\end{table}
\begin{table}[htb]
\small
\caption{$A_N$ vs $p_T$ for $K^+$ and $K^-$ -mesons.}
\begin{center}
\begin{tabular}{|c|c|c|c|c|}    \hline
             &                   &                    \\[0.01cm] 
$p_T$ (GeV/c)& $A_{N}^{K^{+}}$ & $A_{N}^{K^{-}}$  \\[0.3cm] \hline
 0.661 & -0.60 $\pm$ 0.46  & +0.65 $\pm$ 0.42  \\[0.3cm] \hline
 0.912 & +0.01 $\pm$ 0.15  & +0.15 $\pm$ 0.16  \\[0.3cm] \hline
 1.131 & -0.074$\pm$ 0.049 & -0.046$\pm$ 0.051 \\[0.3cm] \hline
 1.368 & +0.011$\pm$ 0.029 & +0.004$\pm$ 0.039 \\[0.3cm] \hline
 1.613 & +0.005$\pm$ 0.030 & -0.000$\pm$ 0.051 \\[0.3cm] \hline
 1.861 & -0.016$\pm$ 0.038 & +0.034$\pm$ 0.077 \\[0.3cm] \hline
 2.110 & +0.135$\pm$ 0.052 & +0.06 $\pm$ 0.12  \\[0.3cm] \hline
 2.360 & +0.056$\pm$ 0.075 & +0.05 $\pm$ 0.16  \\[0.3cm] \hline
 2.613 & +0.03 $\pm$ 0.11  & +0.40 $\pm$ 0.19  \\[0.3cm] \hline
 2.864 & +0.10 $\pm$ 0.15  & +0.36 $\pm$ 0.23  \\[0.3cm] \hline
 3.118 & +0.33 $\pm$ 0.18  & +0.00 $\pm$ 0.27  \\[0.3cm] \hline
 3.368 & +0.28 $\pm$ 0.23  & +0.65 $\pm$ 0.29  \\[0.3cm] \hline
\end{tabular}
\end{center}
\label{Tab2}
\end{table}
\begin{table}[htb]
\small
\caption{$A_N$ vs $p_T$ for protons and antiprotons.}
\begin{center}
\begin{tabular}{|c|c|c|c|c|}    \hline
             &             &                    \\[0.01cm] 
$p_T$ (GeV/c)& $A_{N}^{p}$ & $A_{N}^{\bar{p}}$  \\[0.3cm] \hline
 0.912 & -0.024$\pm$ 0.082 & +0.44 $\pm$ 0.20  \\[0.3cm] \hline
 1.131 & -0.064$\pm$ 0.029 & +0.017$\pm$ 0.067 \\[0.3cm] \hline
 1.368 & -0.047$\pm$ 0.020 & -0.023$\pm$ 0.068 \\[0.3cm] \hline
 1.613 & -0.071$\pm$ 0.025 & +0.09 $\pm$ 0.12  \\[0.3cm] \hline
 1.861 & -0.050$\pm$ 0.020 & -0.17 $\pm$ 0.17  \\[0.3cm] \hline
 2.110 & -0.013$\pm$ 0.023 & +0.08 $\pm$ 0.22  \\[0.3cm] \hline
 2.360 & -0.066$\pm$ 0.030 & +0.19 $\pm$ 0.29  \\[0.3cm] \hline
 2.613 & -0.085$\pm$ 0.043 & +0.02 $\pm$ 0.34  \\[0.3cm] \hline
 2.864 & -0.064$\pm$ 0.060 & +1.06 $\pm$ 0.37  \\[0.3cm] \hline
 3.118 & -0.009$\pm$ 0.085 & +0.44 $\pm$ 0.41  \\[0.3cm] \hline
 3.368 & +0.05 $\pm$ 0.11  & -0.21 $\pm$ 0.47  \\[0.3cm] \hline
\end{tabular}
\end{center}
\label{Tab3}
\end{table}
  Asymmetry for $\pi^{+}$ is shown in Fig.3, as a function of $x_{T}$.
The lines in the figure represent fit by expression:
\begin{equation}
   A_{N} = A_{0}(x_{T}-X_{0}), \quad
\label{EQ1}
\end{equation}    
where slope ($A_{0}$) and sign change point ($X_{0}$) are free parameters. 

The  bin width for $p_{T}$ is 0.25 GeV/c and mean $p_{T}$ varies
from 0.66 to 3.37 GeV/c.
The plotted values of $x_{T}$ correspond to the mean values within the bins.
Since the laboratory angle of the FODS-2 spectrometer is $9^{o}$, mean
values of corresponding $x_{F}$ at 40 GeV beam energy
are not exactly zero and are rising from 0.02 to 0.10 with the rise of $p_{T}$.
  Asymmetry for $\pi^{+}$ is rising with $x_{T}$ and changes sign near 
$x_{T} = 0.37$.

The data for the BNL experiment~\cite{SAR}  are  also shown in Fig.3
for comparison.
It is seen from Fig.3 that $A_{N}$ changes sign at the same
$x_{T}$ for three different beam energies (13.3, 18.5 and 40 GeV). 
The cut $x_{F}\le 0.18$ is used for the BNL data to compare 
with this experiment.

  Asymmetry for $\pi^{-}$  is shown in Fig.4, where BNL data are
also shown for comparison.
  Asymmetry for $\pi^{-}$ is  negative
and it is rising in absolute value with $x_{T}$ increase.
 $A_{N}$ at the BNL
energies is close to zero with some indication for positive slope $A_{0}$.

  Asymmetry for $K^{+}$ (Fig.5)
is rising with $x_{T}$ similar to 
$\pi^{+}$ data and changes sign at  $x_{T}$ near 0.37.

Asymmetries for $K^{-}$ and antiprotons (Fig.5)
 have an indication of the rise with $x_{T}$ similar to
$K^{+}$ data but statistical errors are too high to make this
conclusion final. The last three points for \=p from Table~\ref{Tab3}
are combined in Fig.5 to decrese error.

The comparison of $A_{N}$ for protons with the BNL data are shown
in Fig.6, as a function of $x_{T}$.
For both experiments  $A_{N}$
does not depend on $x_{T}$ and at 40 GeV
the mean value of $A_{N}$ is $-0.050 \pm 0.009$.
The mean value of $A_{N}$ within explored range
of $x_{T}$ is shown  in Fig.7
 for three energies as a function of 
$\sqrt{s}$.
It is seen from Fig.7, that the mean value of $A_{N}$ is
well described  by a linear function of $\sqrt{s}$:
\begin{equation}
A_{N}=(0.0113 \pm 0.0027)(4.34 \pm 0.45-\sqrt{s}/GeV). \quad
\end{equation}
This result, if it is confirmed by other experiments, can be used
to measure beam polarization in a wide range of energies.

  As was mentioned above asymmetry for particle ratios, which  under the first
approximation equals the difference of asymmetries for two types of
particles, 
is not sensitive to beam monitoring and thus has less systematics.
Predictions of models in many cases are also better defined for
particle ratios than for each particle separately.
  As an example of asymmetries for particle ratios Fig.8
shows $A_{N}$
for $K^{+}/\pi^{+}$ and $K^{-}/\pi^{-}$ ratios.
For $K^{+}/\pi^{+}$-ratio
$A_{N}$ is consistent
with zero, as could be expected, since the fragmentation of
scattered valence $u$-quark is
a main production source of both mesons~\cite{FODSR}.
Asymmetry for $K^{-}/\pi^{-}$ is
increasing with $x_{T}$ and it is positive for $x_{T} \ge$ 0.28.

Fit parameters for the data, shown in Figs. 3-8, are presented in
Table~\ref{Tab4}.
As is seen from Table~\ref{Tab4} there is a significant
slope ($A_{0}$)
in dependence on $A_{N}$ vs $x_{T}$ (\ref{EQ1})
for all particles except protons.
The sign of the slope is positive
for particles containing valence $u$-quark of colliding hadrons
 (for $\pi^{+},K^{+}$),
or sea quark (for $K^{-}$,\=p). For particles containing 
valence $d$-quark from colliding hadrons (for $\pi^{-}$) it is negative.
\begin{table}[htb]
\small
\caption{Parameters $X_{0}$ and $A_{0}$ of equation (4) as a
  function of energy.}
\begin{center}
\begin{tabular}{|c|c|c|c|c|c|}    \hline
Ref.   &hadron & E (GeV)  &$X_{0}$         &$A_{0}$           \\[0.3cm] \hline
FODS-2 &$\pi^+$& 40.0     &$0.37\pm0.02$   &$ 0.33\pm0.08$    \\[0.3cm] \hline
FODS-2 &$\pi^-$& 40.0     &$0.06\pm0.19$   &$-0.08\pm0.05$    \\[0.3cm] \hline
FODS-2 &$\ K^+$& 40.0     &$0.35\pm0.04$   &$ 0.46\pm0.16$    \\[0.3cm] \hline
FODS-2 &$\ K^-$& 40.0     &$0.32\pm0.04$   &$ 0.59\pm0.25$    \\[0.3cm] \hline
FODS-2 & \=p   & 40.0     &$0.24\pm0.12$   &$ 0.43\pm0.39$    \\[0.3cm] \hline
 [3]   &$\pi^+$& 13.3     &$0.33\pm0.04$   &$ 0.33\pm0.11$    \\[0.3cm] \hline
 [3]   &$\pi^+$& 18.5     &$0.37\pm0.02$   &$ 0.58\pm0.14$    \\[0.3cm] \hline
 [3]   &$\pi^-$& 13.3     &$0.018\pm0.024$ &$ 0.037\pm0.019$  \\[0.3cm] \hline
 [3]   &$\pi^-$& 18.5     &$0.007\pm0.005$ &$ 0.025\pm0.026$  \\[0.3cm] \hline
FODS-2 &$K^{+}/\pi^+$ &40.0&$0.27\pm0.23$   &$ 0.11\pm0.17$   \\[0.3cm] \hline
FODS-2 &$K^{-}/\pi^-$ &40.0&$0.29\pm0.04$   &$ 0.70\pm0.26$   \\[0.3cm] \hline
\end{tabular}
\end{center}
\label{Tab4}
\end{table}

  As was shown above, the asymmetry dependence on $x_{T}$ can be
described by linear function (\ref{EQ1}). To study the dependence of
parameters $A_{0}$ and $X_{0}$ of (\ref{EQ1}) on energy we fitted $x_{T}$
dependence of asymmetry for several experiments in a wide range
of energies~\cite{SAR},~\cite{PROZ},~\cite{ANTIL},~\cite{Adams}.
Since in some experiments only $\pi^{0}$-mesons have been detected,
their asymmetry is
 compared below with mean $A_{N}$ for $\pi^{+}$ and $\pi^{-}$-mesons,
averaged according to a parton model
with weights proportional to their production cross sections ~\cite{Trosh}:
\begin{equation}
 A_{N}(\pi^{o})={ {A_{N}(\pi^{+})\cdot R(x_{T}) + A_{N}(\pi^{-})}
 \over {R(x_{T}) + 1} }, \quad
\label{R}
\end{equation}
where $R(x_{T})$ is a ratio of the production cross sections
for  $\pi^{+}$ and $\pi^{-}$ -mesons.
The resulting dependences of $X_{0}$ and $A_{0}$ on
$\sqrt{s}$ are shown in Figs.9-10,
respectively.
The dependence of $X_{0}$
on $\sqrt{s}$ is consistent with constant value of 
$X_{0} = 0.374 \pm 0.009$, while $A_{0}$ decreases with the rise of energy:
\begin{equation}
 A_{0} = (0.029\pm0.005)*(21\pm2 -\sqrt{s}/GeV), \quad 
\end{equation}
which means the decrease of spin effects with the rise of energy.
Scaling for $X_{0}$, observed in different experiments, proves that we
have no significant systematics for asymmetry in this experiment.
One point in Fig.10,
corresponding to the PROZA-M experiment result is much higher
than other points, which could be related
to different beam ($\pi^{-}$) for that experiment~\cite{PROZ}.
Of course, the exact dependence of $A_{0}$ on energy is unavailable
and the linear expression is just the simplest one.

   It is interesting to compare the observed scaling for
 parameter $X_{0}$ with predictions of theoretical models. Predictions
for asymmetry in the central region for a wide range of energies have been 
recently made in the framework of QCD~\cite{Trosh}. The main role in this
model belongs to the orbital angular momentum of the quark-antiquark cloud
in the internal structure of constituent quarks. For polarization of the
constituent quarks SU(6)-model values $P_{U}$ = 2/3 and $P_{D}$ = -1/3 
are used. Predictions of asymmetry
for $\pi^+$ and $\pi^-$-mesons have been made for 70, 200 and 800 GeV.
These predictions along with their extrapolation to 40 GeV are shown in
Fig.11 as the functions of $x_{T}$.  It is seen from
Fig.11, that the value of $x_{T}$, where
$A_{N}=0$ is decreasing with the energy rise (which corresponds to constant
$p_{T}=0.75$ GeV/c), contrary
to the experimental data where it takes place at constant $x_{T}$ around 0.37.

The   comparison of data and model predictions for 40 GeV
is shown in Fig.12.
The 40 GeV energy model predictions are in a   qualitative agreement with
the results of this experiment. The model predicts the rise of $A_{N}$ with
$x_{T}$, the opposite sign of $A_{N}$ for $\pi^+$ and $\pi^-$, and half as 
much the absolute value of asymmetry for $\pi^-$ against
for $\pi^+$. 
The model predicts the increase for $A_{0}$ and the decrease for $X_{0}$ with
 the rise of energy. That is in some contradiction with the results shown in 
Figs.9-10 for $\pi^0$-mesons.
Asymmetry
for $\pi^0$-meson at 200 GeV  is very small~\cite{Adams}, which is possible
to explain if $P_{U}=-P_{D}$ and hence
$A_{N}^{\pi+}$ = -$A_{N}^{\pi-}$, which is not seen at 40 GeV energy and
below it.

\section {Conclusion}

  The first asymmetry measurements have been performed for $\pi^{\pm}$,
$K^{\pm}$, p and \=p with 40 GeV/c IHEP polarized proton beam.
Asymmetries for $K^{\pm}$ and \=p have never been measured before.
Asymmetries  for $\pi^{\pm}$,
$K^{\pm}$,  \=p show approximately linear dependence on $x_{T}$ and
change the sign near 0.37. The slope sign of this dependence is negative
for $\pi^{-}$ and positive for other hadrons, except protons.
For protons the negative asymmetry, independent
of $x_{T}$ has been found, which grows with energy in absolute value as
the comparison with the other experiment indicates.
   
The   SU(6) model predicts asymmetry that is in a qualitative agreement with
our data for $\pi$-mesons.

A significant increase in statistics is
possible (by a factor around 40) in future runs which will allow us to make
definite conclusion about asymmetries in a wide range of $x_{T}$ and $x_{F}$.

We gratefully acknowledge the assistance and support of the IHEP staff and
directorate.
The research was supported in part by the Russian Fund
for Fundamental Research.


%
\newpage
\listoffigures
%
\newpage
\begin{figure}
\centerline{\epsfig{file=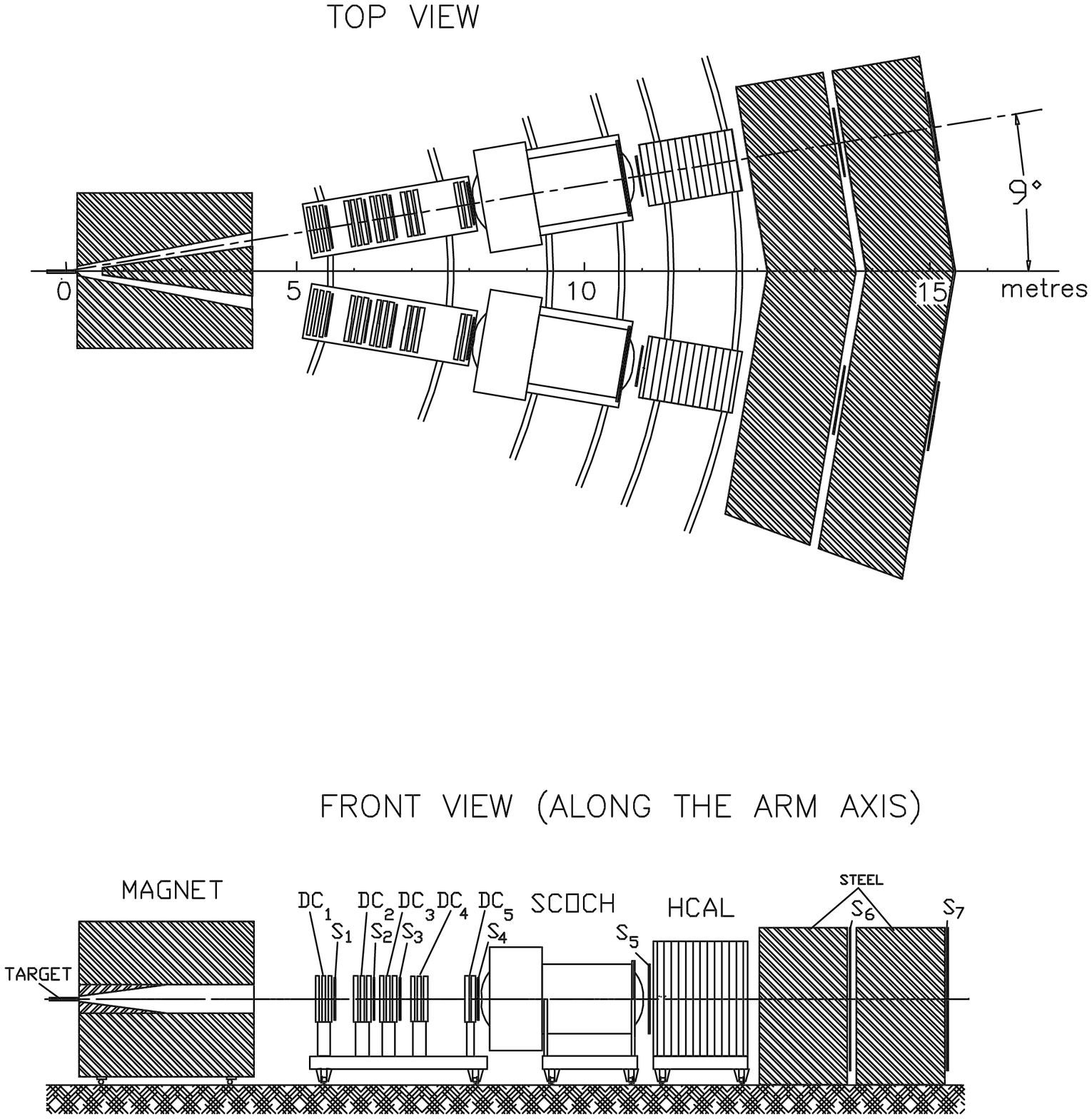,width=14cm}}
\caption {The schematic layout of
FODS-2 spectrometer.}            
\label{FODS}
\end{figure}
\clearpage
\begin{figure}
\centerline{\epsfig{file=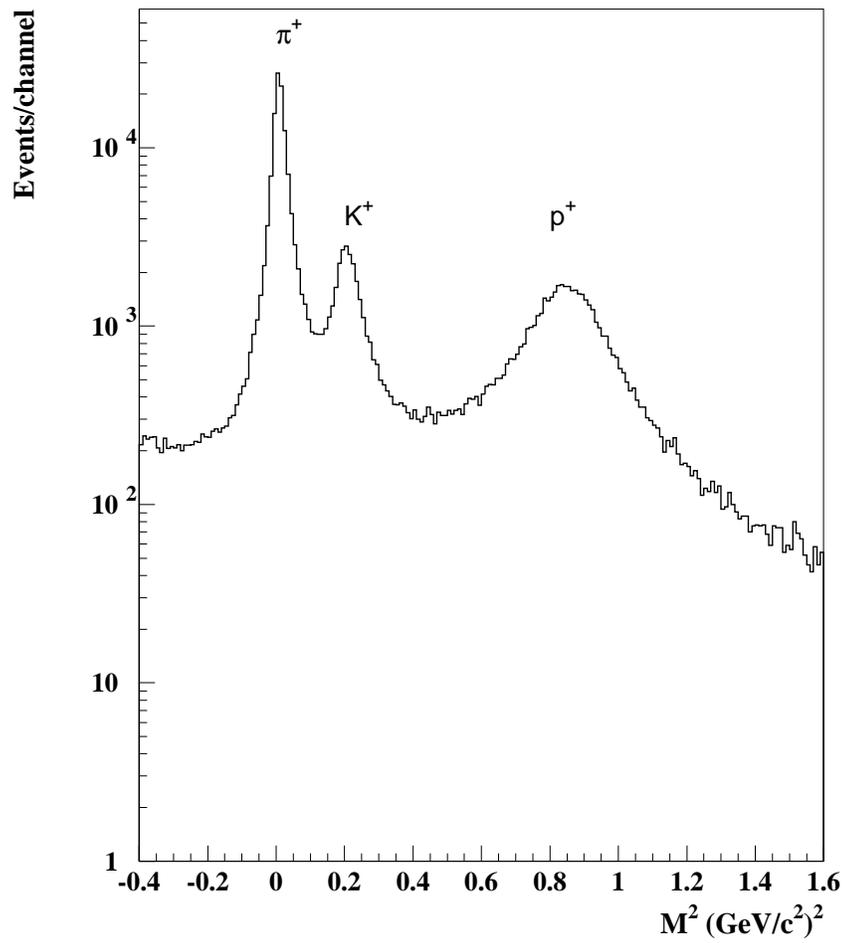,width=14cm}}
\caption {Mass spectrum of 
hadrons measured by SCOCH detector.}
\label{MASS}
\end{figure}
\begin{figure}
\centerline{\epsfig{file=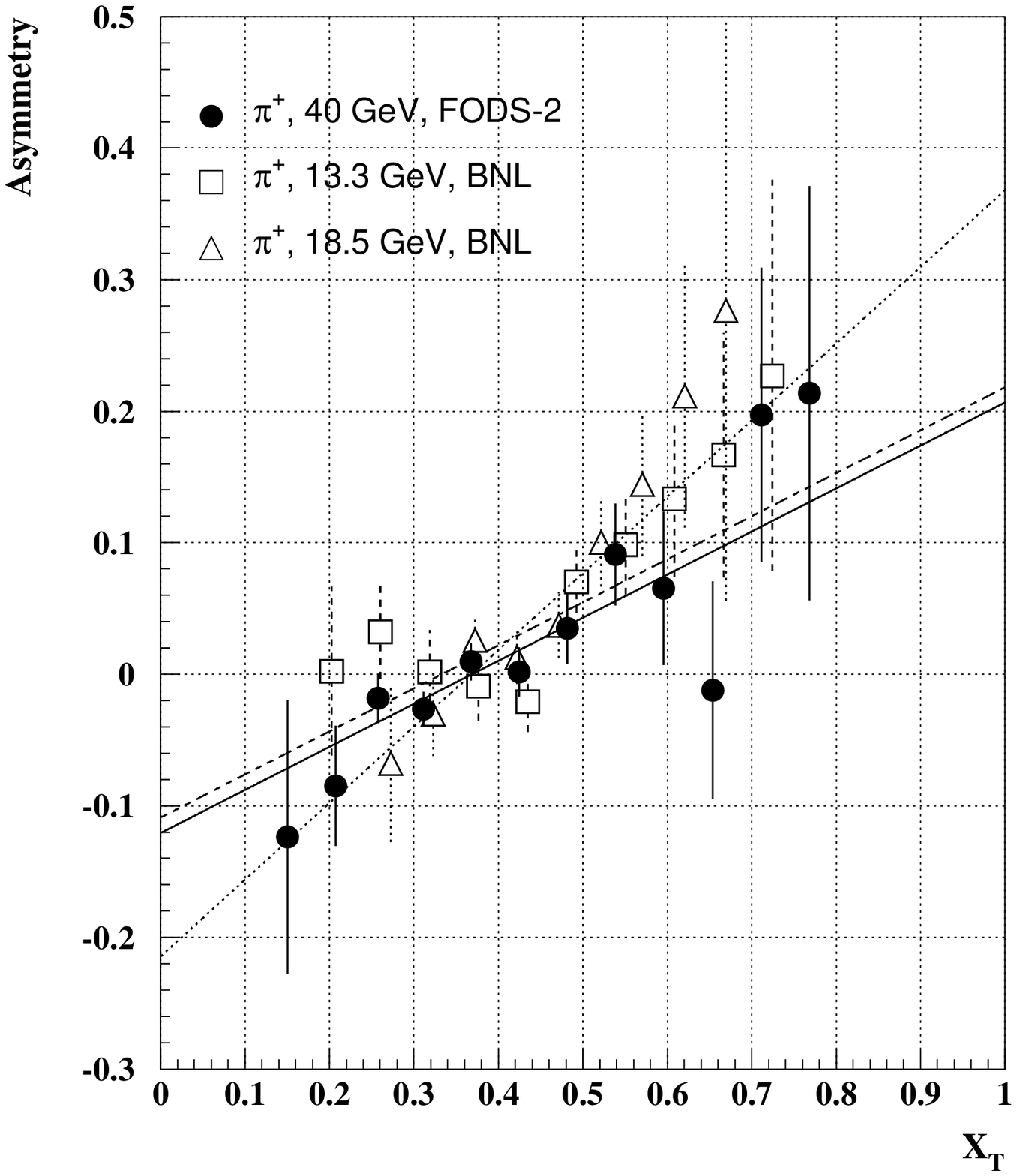,width=14cm}}
\caption {Comparison of $A_{N}$ vs $x_{T}$ for
$\pi^{+}$ -mesons at 40 GeV, 18.5 and 13.3 GeV [3].
Solid line shows fit (5) for 40 GeV, dashed line - for 13.3 GeV,
and dotted line - for 18.5 GeV.}
\label{BNLP}
\end{figure}
\begin{figure}
\centerline{\epsfig{file=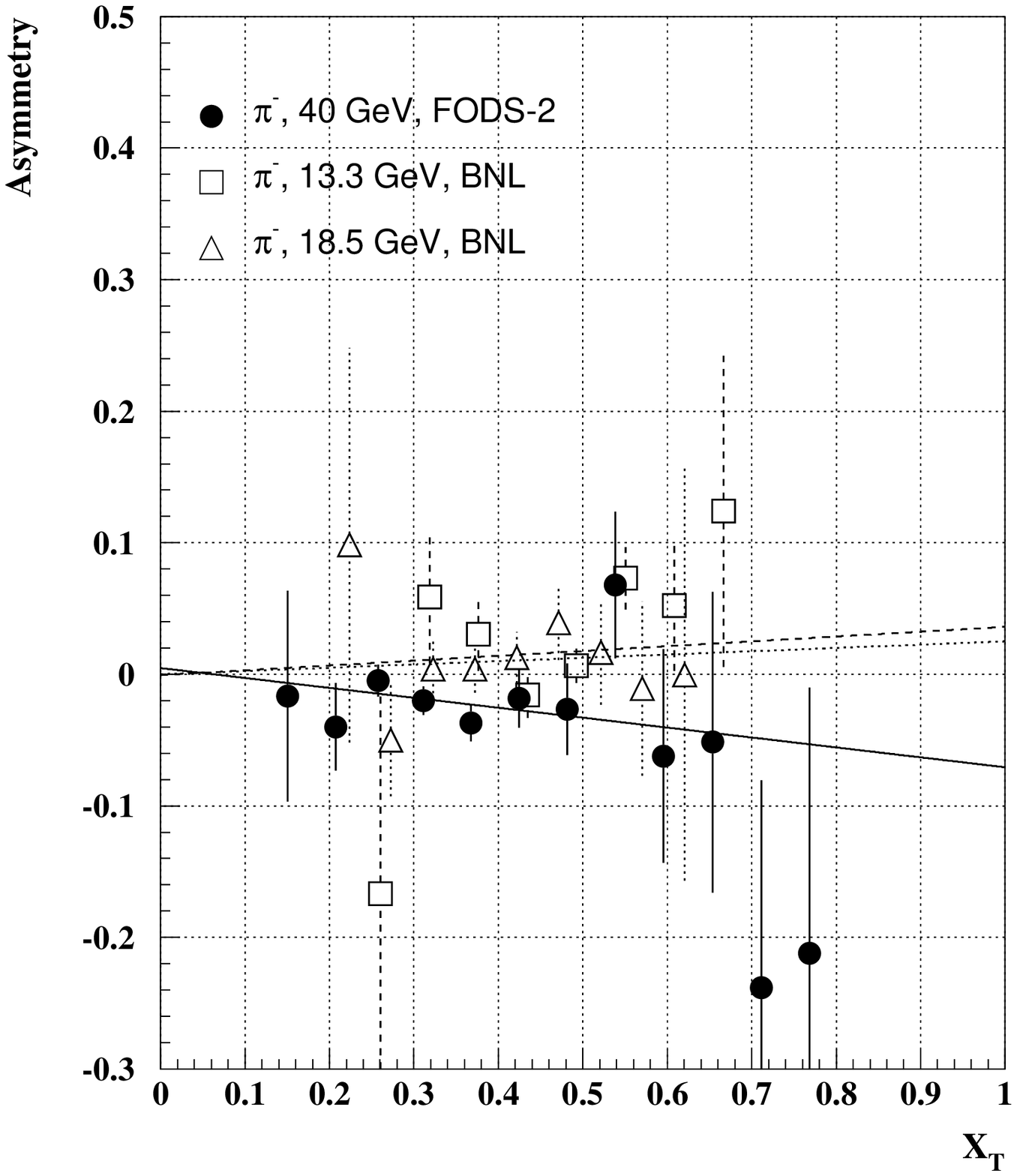,width=14cm}}
\caption {Comparison of $A_{N}$ vs $x_{T}$ for
$\pi^{-}$ -mesons at 40 GeV, 18.5 and 13.3 GeV [3].
Solid line shows fit (5) for 40 GeV, dashed line - for 13.3 GeV,
and dotted line - for 18.5 GeV.}
\label{BNLM}
\end{figure}
\begin{figure}
\centerline{\epsfig{file=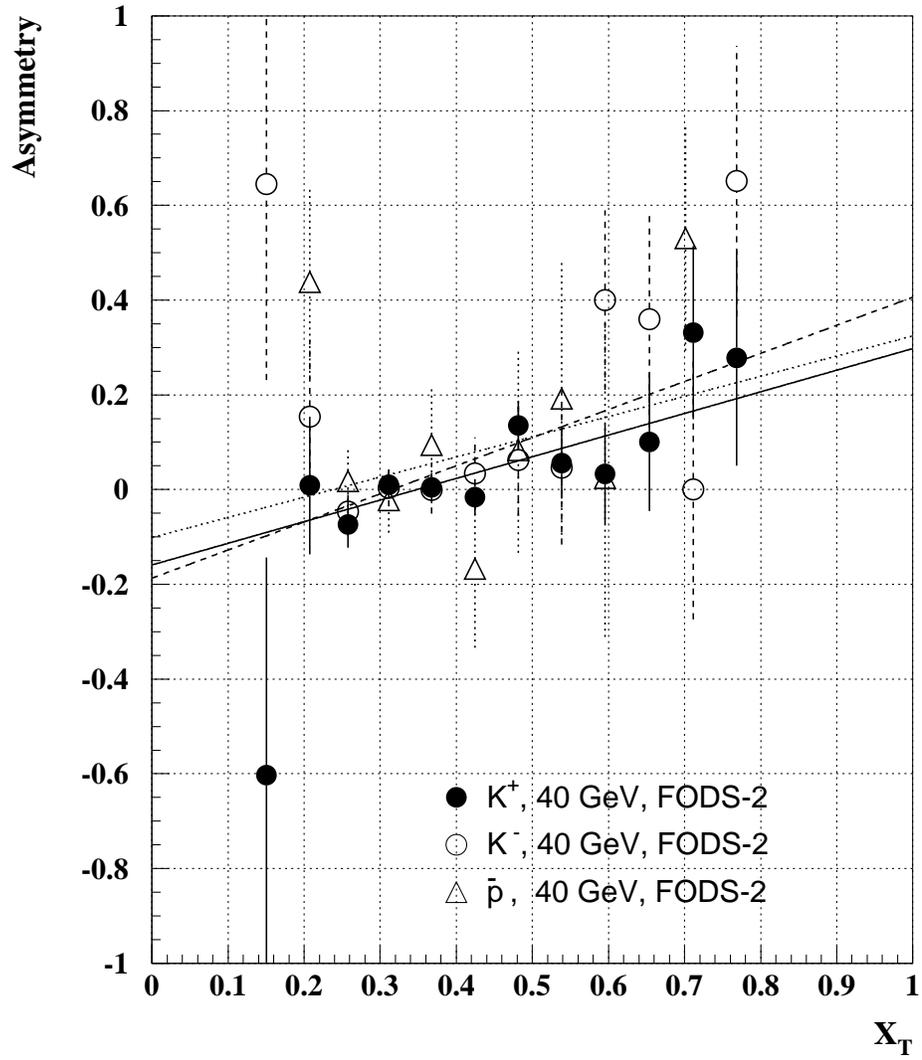,width=14cm}}
\caption {Single-spin asymmetry vs $x_{T}$ for
$K^{+}$, $K^{-}$ -mesons and antiprotons.
Solid line shows fit (5) for $K^{+}$, dashed line - for $K^{-}$,
and dotted line - for antiprotons.}
\label{KMES}
\end{figure}
\begin{figure}
\centerline{\epsfig{file=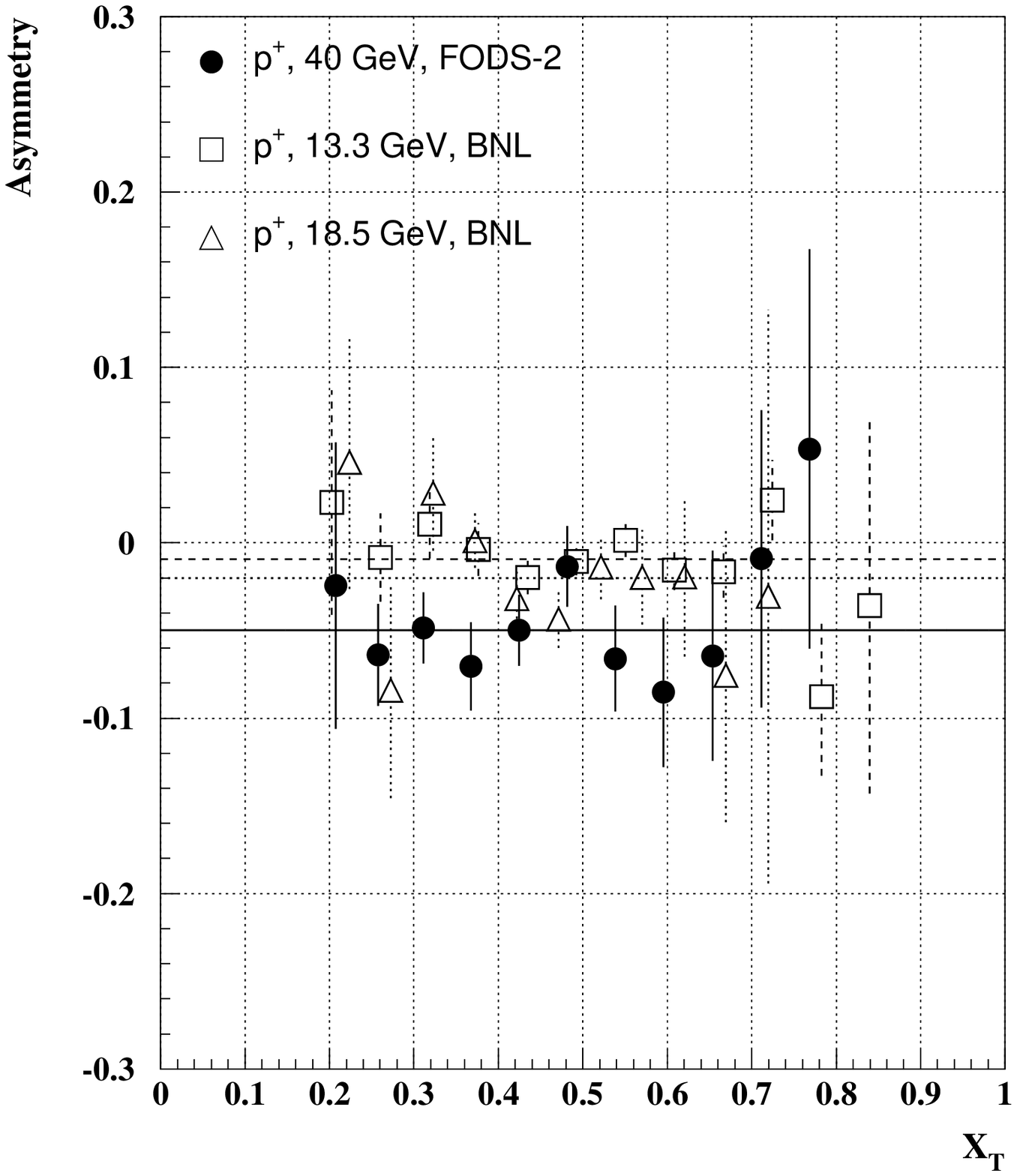,width=14cm}}
\caption {Comparison of $A_{N}$ vs $x_{T}$ for
protons at 40 GeV, 18.5 and 13.3 GeV [3].
Solid line shows fit (5) for 40 GeV, dashed line - for 13.3 GeV,
and dotted line - for 18.5 GeV.}
\label{BNLPR}
\end{figure}
\begin{figure}
\centerline{\epsfig{file=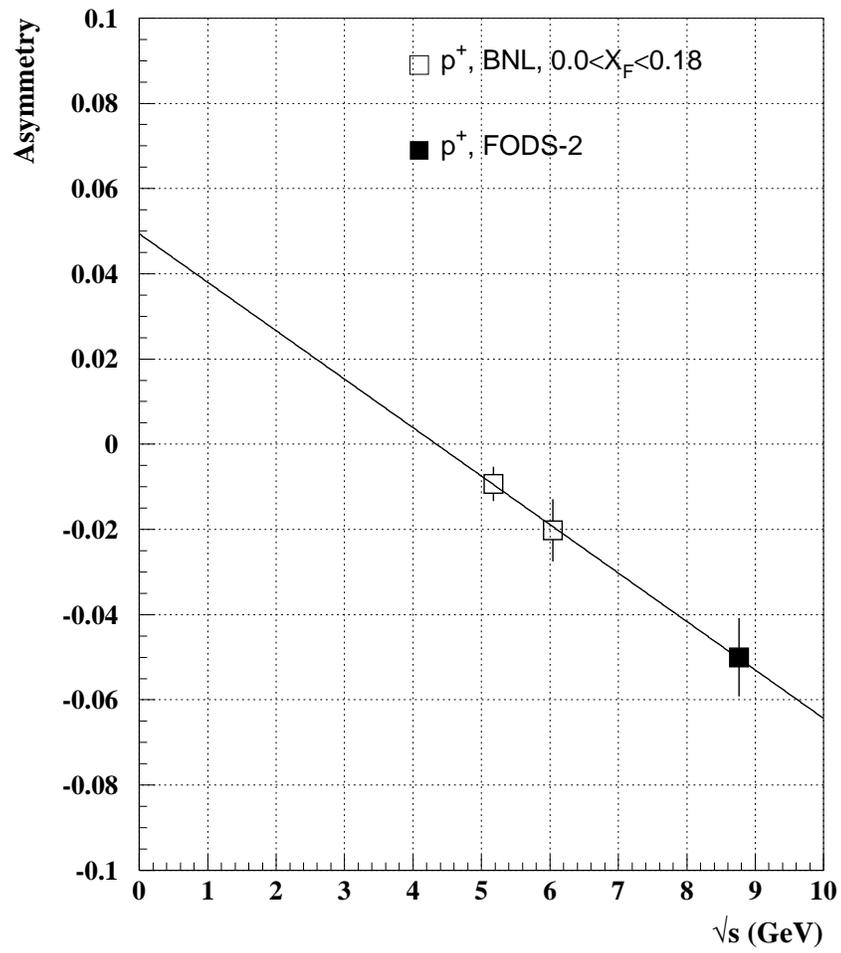,width=14cm}}
\caption {$A_{N}$ vs CM collision energy for
protons for FODS-2 and BNL [3] data.}
\label{BNLPRE}
\end{figure}
\begin{figure}
\centerline{\epsfig{file=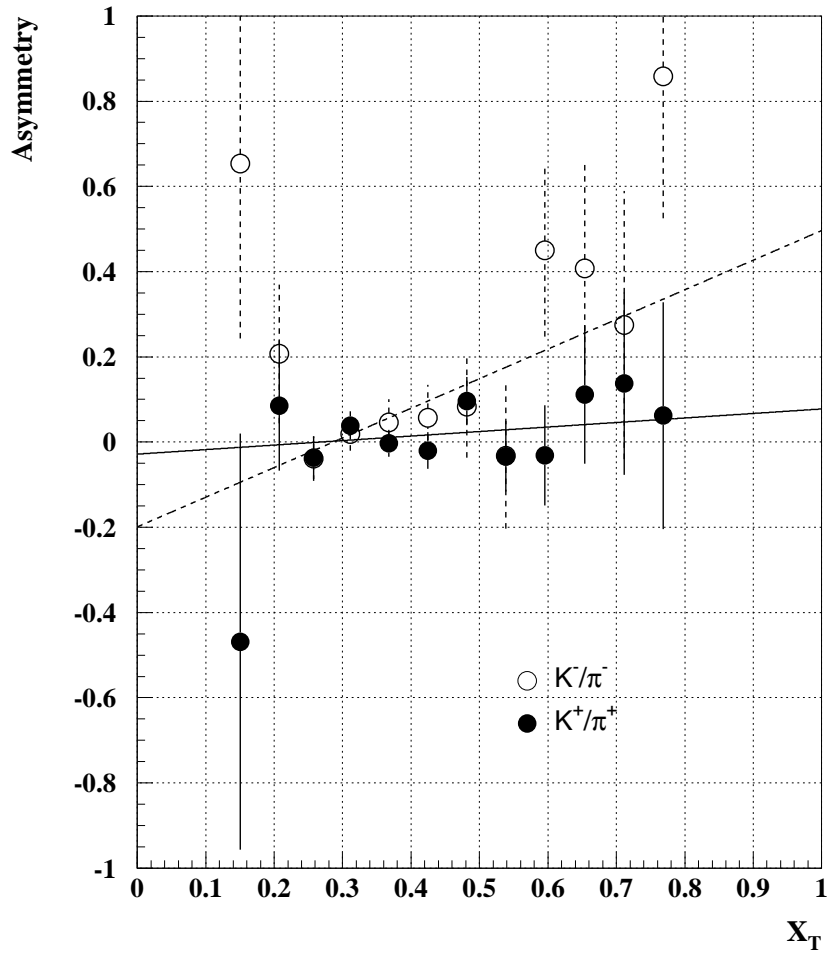,width=14cm}}
\caption {Asymmetry vs $x_{T}$ for particle ratios
$K^{+}/\pi^{+}$ and $K^{-}/\pi^{-}$.
Solid line shows fit (5) for $K^{+}/\pi^{+}$, 
dashed line - for $K^{-}/\pi^{-}$.}
\label{KPI}
\end{figure}
\begin{figure}
\centerline{\epsfig{file=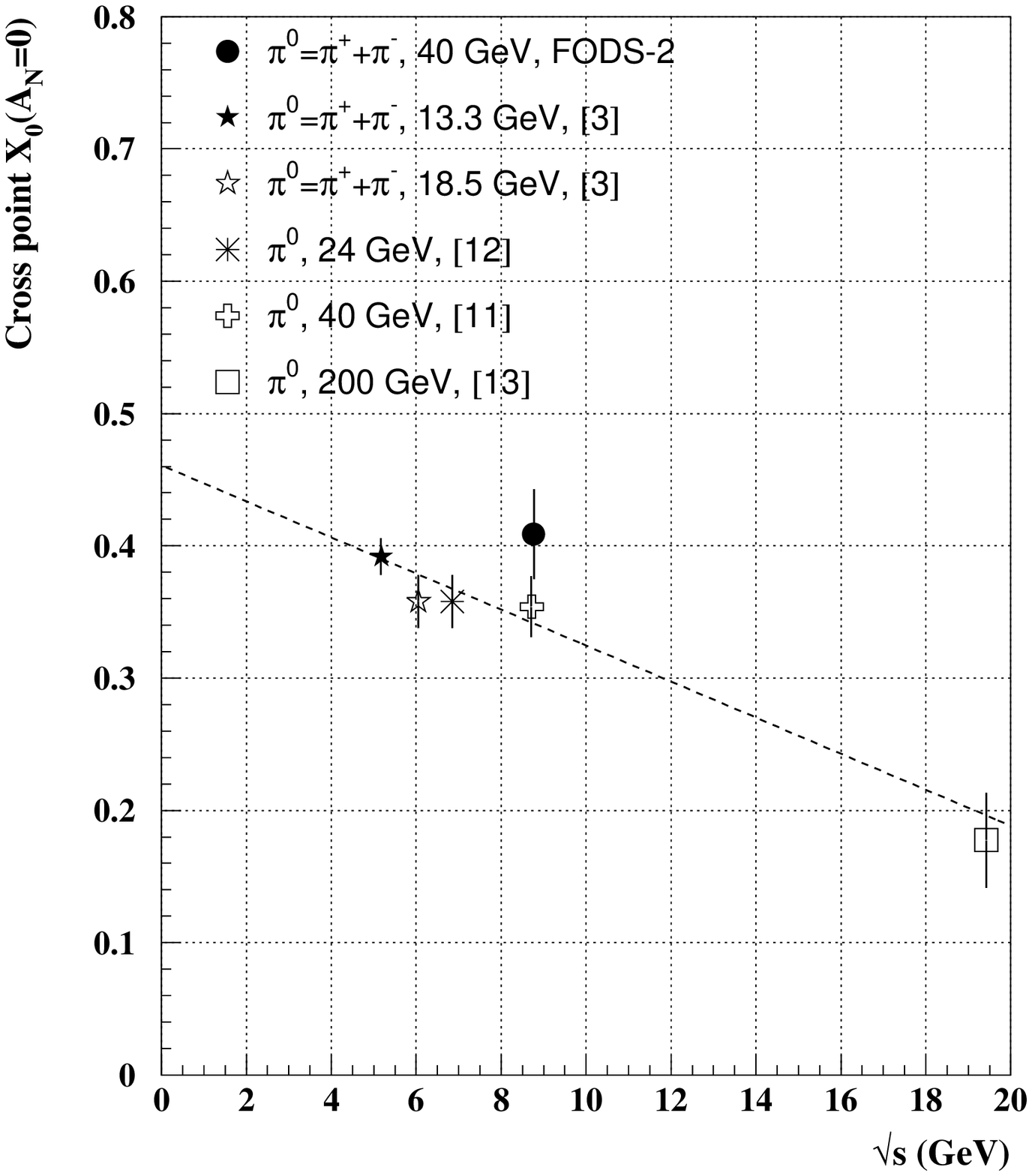,width=14cm}}
\caption {Dependence of parameter $X_{0}$ of eq. (5) vs CM collision  
energy for $\pi^{0}$ [11,12,13] and $\pi^{+}+\pi^{-}$ [3], FODS-2.
Asymmetries for $\pi^{+}$ and $\pi^{-}$ are added with weights
proportional to the production cross sections for these mesons
following (7).}
\label{X0VSE}
\end{figure}
\begin{figure}
\centerline{\epsfig{file=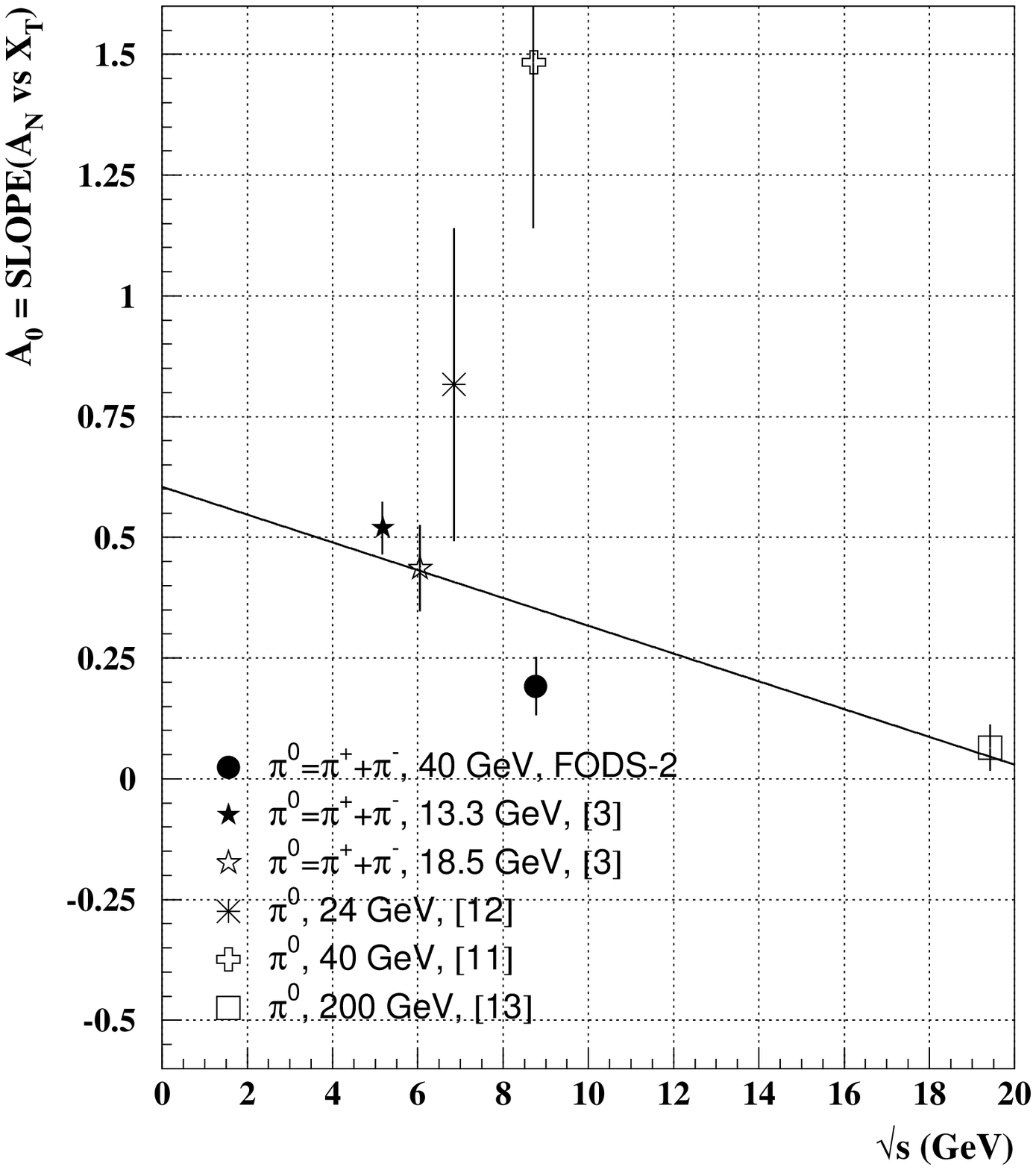,width=14cm}}
\caption {Dependence of parameter $A_{0}$ of eq. (5) vs CM collision 
energy for $\pi^0$ [11,12,13] and $\pi^{+}+\pi^{-}$ [3], FODS-2.
Asymmetries for $\pi^{+}$ and $\pi^{-}$ are added with weights
proportional to the production cross sections for these mesons
following (7).}
\label{A0VSE}
\end{figure}
\begin{figure}
\centerline{\epsfig{file=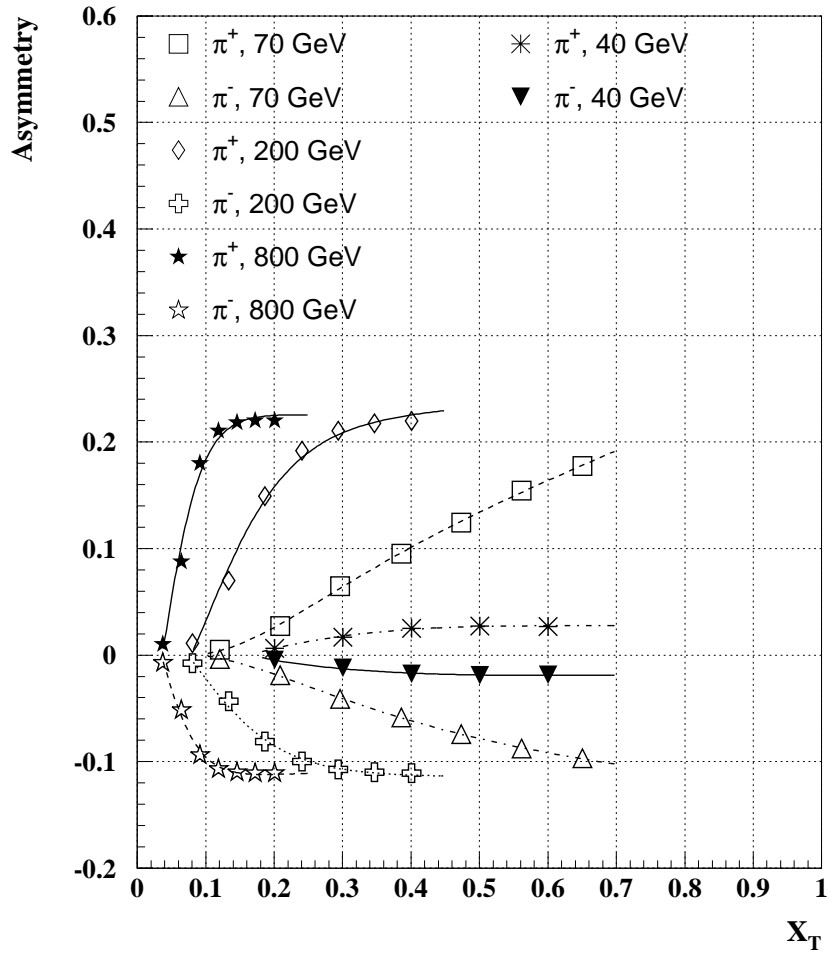,width=14cm}}
\caption {SU(6) model predictions of $A_{N}$ vs $x_{T}$ for
$\pi^{+}$ and $\pi^-$ -mesons at different energies [14].}
\label{SU6XT}
\end{figure}
\begin{figure}
\centerline{\epsfig{file=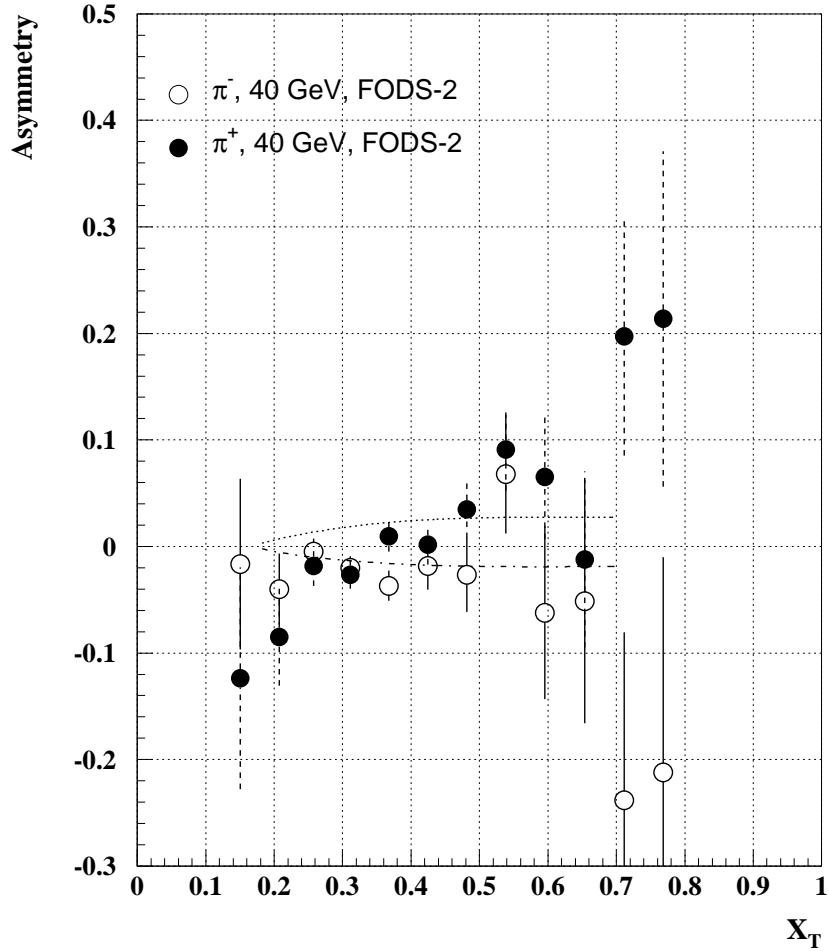,width=14cm}}
\caption {Comparison of data and SU(6) model [14] predictions at 40 GeV.
 Data ($A_{N}$ vs $x_{T}$) are shown for
$\pi^+$ and $\pi^-$ -mesons. Predictions are shown by dotted curve
($\pi^+$) and by dash-dotted curve ($\pi^-$).}
\label{SU6DAT}
\end{figure}
\end{document}